\documentclass[12pt]{article}
\usepackage[square,numbers,sort&compress]{natbib}
\usepackage[pdftex]{graphicx}
\usepackage{authblk}
\title{Integral equation study of effective attraction between like-charged particles mediated by cations: Comparison between IPY2 and HNC closures}
\begin{document}
\author[1]{Michika Takeda}
\affil[1]{Department of Chemistry,
            Kyushu University,
            Fukuoka,
            812-0395,
            Japan}
\author[2]{Kotetsu Maruyama}

\affil[2]{Matsuyama Minami High School, 
            Ehime,
            790-0023,
            Japan}

\affil[3]{Department of Physics,
            Ehime University,
            Ehime,
            790-8577,
            Japan}
\author[1]{Ryo Akiyama \thanks{Corresponding author. \it Email address\rm:\ rakiyama@chem.kyushu-univ.jp}}
\author[3]{Tatsuhiko Miyata \thanks{Corresponding author. \it Email address\rm:\ miyata.tatsuhiko.mf@ehime-u.ac.jp}}
\maketitle
\begin{abstract}
Effective interactions between like-charged particles immersed in an electrolyte solution were calculated using two integral equation theories, hypernetted-chain (HNC)-Ornstein--Zernike (OZ) and ionic Percus--Yevick 2 (IPY2)-OZ. When the HNC-OZ theory was adopted, the electrolyte concentration dependence of the effective interaction showed a reentrant behavior. By contrast, the IPY2-OZ theory did not indicate the behavior. Monte Carlo simulations were performed for one of the model systems, and the results agreed qualitatively with those calculated using the HNC-OZ theory.
\end{abstract}

\section{Introduction}

Acidic proteins in an electrolyte solution attract each other under certain conditions \cite{Zhang2008,zhang2010,Zhang2012,Zhang2017_2,Zhang2021}.
The condensation behavior depends on the electrolyte concentration. When the electrolyte concentration is low, the proteins repel each other; 
but if the electrolyte concentration increases to some mM level, condensation appears, and with further concentration increase, the protein aggregations redissolve.
This reentrant condensation behavior has been reported by Zhang et al. when the electrolyte solution has multivalent cations, such as $Y^{3+}$ \cite{Zhang2008,zhang2010,Zhang2012,Zhang2017_2,Zhang2021}.

To study effective interaction, theoretical approaches such as the integral equation theories (IETs) were examined \cite{Guldbrand1984,Svensson1984,moreira2002simulations,Wu2005ion,Valleau1991,Bratko1998,Bratko1999,WuBratko2000,Ikeda2014,Li2017,Abrikosov2017,Bratko2018,Nguyen2000,Chen2004,Zhang2014,Ounchic2015,oosawa1971polyelectrolytes,manning_1978,Wilson1979,Sumi2009,KJELLANDER1984,Kjellander1988,Kjellander1988_2,Bratko1986,Kjellander1990,Kinoshita1996,Bratko1986_3,Bratko1986_2,patey1980,Frusawa_2009,Akiyama2011,AkiyamaSakata2011,FUJIHARA2014,suematsu2018,Suematsu2021,Yoshimori2020,babu1994new}. 
Some results indicated an effective attraction between like-charged particles. 
Despite the high costs due to the strong direct attraction between anions and cations, some researchers performed simulation studies. The results also showed the existence of effective attraction \cite{Valleau1991,Bratko1998,Bratko1999,WuBratko2000,Ikeda2014,Li2017,Abrikosov2017,Bratko2018}. 
By contrast, the reentrant behaviors were not shown theoretically until 2011. 
A study using the hypernetted-chain (HNC)-Ornstein--Zernike (OZ) theory showed the reentrant behaviors in effective interaction \cite{Akiyama2011,AkiyamaSakata2011}, and some studies followed to describe the reentrant behavior in effective interaction \cite{FUJIHARA2014,suematsu2018,Suematsu2021}. 
Finally, a thermodynamic perturbation approach with effective interactions showed the reentrant condensations \cite{Yoshimori2020}.

The results obtained from IET are affected by the closure approximation employed. 
Under each condition, the HNC approximation was shown to be valid for monovalent ionic systems such as the 1:1 electrolyte solution \cite{Hansen1975,Friedman1981electrolyte,hansen1990theory}. 
However, it was also shown that for highly associating ionic systems such as 2:2 electrolytes, the HNC results become unsatisfactory qualitatively. 
To overcome such obstacles, the ionic Percus--Yevick 2 (IPY2) closure relation was proposed\cite{babu1994new}. 
The results obtained under the IPY2 approximation reproduced the Monte Carlo (MC) simulation data for 2:2 electrolyte solutions well \cite{babu1994new}. In protein condensation, which is the main target of this study, a mixture of multivalent and monovalent ions is typically involved as an experimental condition.
For such a case, it is still unclear which theory is more appropriate between the HNC and IPY2. 

To study the existence of an effective attraction and reentrant behavior, we calculated the effective interaction between like-charged particles immersed in an electrolyte solution. In the present study, the results calculated using the IPY2-OZ and HNC-OZ theories are compared. If the results in the discussion of the adequate closure are contradictory, MC simulation will indicate which one is crucial.

\section{Model and Method}
\subsection{Direct Interaction between Particles}

A charged hard-sphere model was adopted in the present study. The direct interaction between particles is shown as follows:
\begin{equation}
    u_{ij}(r)=\left\{\begin{array}{cc}
    \infty & (r<\sigma_{ij}) \\[1.5ex]
    \displaystyle\frac{Q_{i}Q_{j}}{4\pi\varepsilon_{0}\varepsilon_{r}r} & (r\geq \sigma_{ij}) ,
    \end{array}\right.
\end{equation}
where $u_{ij}(r)$ is the interaction between particles $i$ and $j$ at the distance $r$, and $Q_i$, $\varepsilon_0$, $\varepsilon_r$, and $\sigma_{ij}$ are the charge of particle $i$, the permittivity of vacuum, the relative permittivity of water (78.5), and the contact distance between $i$ and $j$, respectively. The contact distance $\sigma_{ij}$ is $(\sigma_{ii}+\sigma_{jj})/2$, where $\sigma_{ii}$ is the diameter of particle $i$.

\subsection{Model A}
The first model, Model A, was prepared to calculate the effective interaction between macroanions immersed in an electrolyte solution. This model was similar to that used in Ref.\cite{AkiyamaSakata2011}. We deal with this model to compare our previous results, obtained using the HNC closure in Ref.\cite{AkiyamaSakata2011}, with those obtained using the IPY2 closure. In Model A, we adopted an explicit solvent model considering the solvent granularity. Thus, the solvent consists of uncharged hard spheres. The solution also has a 1:1 electrolyte. A pair of macroanions are immersed in the solution. ``A'', ``C'', ``M'', and ``V'' denote the anion, cation, macroanion, and solvent, respectively. The anions, cations, and solvents have the same size. The parameters are shown in Table \ref{tab:1_parameta}. In this model, the total packing fraction is kept at 0.383, which is the packing fraction of water. This total packing fraction is maintained during the electrolyte concentration change. The temperature is 298\,K. This is a simplified model of an aqueous electrolyte solution. We calculate the effective interactions between macroanions using the OZ equation with two closures, namely the HNC and the IPY2 closures.
\begin{table}[b]
    \centering
    \caption{Parameter for the models. Here, $\mathrm{e}$ is elementary charge.}
    \scalebox{0.93}{
    \begin{tabular}{l|c|c|c|c}\hline
        & \multicolumn{2}{c|}{Model A (HNC, IPY2)} & \multicolumn{2}{c}{Model B (HNC, IPY2, MC)} \\
        \hline
        & diameter[\AA] & charge[C] & diameter[\AA] & charge[C] \\
        \hline \hline
        Anion (A) & 2.8 & -e & 3.62 & -e \\
        Cation (C) & 2.8 & e & 1.7 & 2.4\,e \\
        Macroanion (M) & 16.8 & -10\,e & -- & -- \\
        Solvent molecule (V) & 2.8 & 0 & -- & -- \\
        \hline
    \end{tabular}
    }
    \label{tab:1_parameta}
\end{table}

\subsection{Model B}
We prepared the second model, Model B, to compare the radial distribution functions $g(r)$ obtained using the IET and the MC simulations. 
Because the calculation cost for ``Model A'', that was mentioned above, was too expensive, the solvent hard spheres and the macroanions were removed from it. The results calculated using the HNC closure showed that the solvent hard spheres intensify the effective attraction between like-charged particles\cite{suematsu2018,Suematsu2021}. However, the explicit solvent model was not required, and the implicit solvent model also showed the effective attraction and the reentrant behavior in our studies\cite{suematsu2018,Suematsu2021}. Therefore, this model was selected to reduce the computational costs.

In this model, the $g_{AA}(r)$s values are compared. In the relationship of $g(r) = \mathrm{exp}[-\beta W(r)]$, $g(r)$ and $W(r)$ are the radial distribution function and the effective interaction, respectively. Thus, the calculated effective interactions can be compared. By contrast, the differences between the radial distribution functions were more apparent than those between the effective interactions in the important $r$ region. Therefore, we decided to use $g(r)$. 
We constructed this model referring to the previous study\cite{suematsu2018}. 
In this model, $\sigma_{AA}$ is 3.62\,\AA\, and $\sigma_{CC}$ is 1.70\,\AA. 
These values correspond to $\mathrm{Cl^-}$ and $\mathrm{Mg^{2+}}$, whereas $Q_C$ is 2.4\,e. Table \ref{tab:1_parameta} shows the parameters. 
When these parameters are adopted, the radial distribution function between anions can be expected to show the reentrant behavior\cite{suematsu2018}.

\subsection{The IETs to Calculate Model A}
We numerically solve the OZ equation with a closure relation. The OZ equation for a four-component system is shown as
\begin{equation}
    h_{ij}(r)=c_{ij}(r)+\sum_{l=C,A,M,V}\rho_l\int c_{il}(\mathbf{r'})h_{lj}\left(|\mathbf{r}-\mathbf{r'}|\right)d\mathbf{r'},
\end{equation}
where $h_{ij}(r)$, $c_{ij}(r)$, and $\rho_i$ are the total correlation function, the direct correlation function, and the number density of $i$, respectively. Here, $\rho_M$ goes to zero because we are studying the dilution limit of the macroanion. At the dilution limit, we obtain three OZ equations. One is the OZ equation for the bulk electrolyte solution as follows:
\begin{equation}\label{eqn:oz_ij}
    h_{ij}(r)=c_{ij}(r)+\sum_{l=C,A,V}\rho_l\int c_{il}(\mathbf{r'})h_{lj}\left(|\mathbf{r}-\mathbf{r'}|\right)d\mathbf{r'},
\end{equation}
and another is the OZ equation for the macroanion and electrolyte solution, as follows:
\begin{equation}\label{eqn:oz_Mi}
    h_{Mi}(r)=c_{Mi}(r)+\sum_{l=C,A,V}\rho_l\int c_{Ml}(\mathbf{r'})h_{li}\left(|\mathbf{r}-\mathbf{r'}|\right)d\mathbf{r'}.
\end{equation}

At first, the OZ equation for the bulk system is solved with a closure relation. The calculated correlation functions become the input functions for the second OZ equation. Eq. (\ref{eqn:oz_Mi}) with a closure provides the correlation functions $h_{Mj}(r)$, $c_{Mj}(r)$.
The effective interaction  $W_{MM}(r)$ can be calculated using the correlation functions $h_{Mj}(r)$, $c_{Mj}(r)$, and the third OZ equation with a closure. The Fourier transformation is as follows:
\begin{equation}\label{eqn:wavenum}
    T_{MM}(k)=\sum_{l=A,C,V}\rho_lH_{Ml}(k)C_{lM}(k),
\end{equation}
where $T_{MM}(k)$, $H_{Ml}(k)$, and $C_{lM}(k)$ are the Fourier transforms of $t_{MM}(r)(= h_{MM}(r) - c_{MM}(r))$, $h_{Ml}(r)$, and $c_{lM}(r)$, respectively; and $k$ is the wave number. When we obtain $T_{MM}(k)$ (or $t_{MM}(r)$), we can calculate $W_{MM}(r)$ using a closure relation:

Before the explanation of $W_{MM}(r)$, we introduce the closure relations. The closures are expected as follows:
\begin{equation}\label{eqn:closure}
    h_{ij}(r)=\exp\left[-\beta u_{ij}(r)+h_{ij}(r)-c_{ij}(r)+b_{ij}(r)\right]-1,
\end{equation}
where $\beta=1/k_BT$, $k_B$ is the Boltzmann constant; $b_{ij}(r)$ is the bridge function. In the present study, we examined two closure relations: HNC and IPY2. In the HNC, $b(r) = 0$ in Eq.\,(\ref{eqn:closure}). It has been used to study electrolyte solutions\cite{KJELLANDER1984,Kjellander1988,Kjellander1988_2,Bratko1986,Kjellander1990,Kinoshita1996,Bratko1986_3,Bratko1986_2,patey1980,Akiyama2011,AkiyamaSakata2011,FUJIHARA2014,suematsu2018,Suematsu2021}. The equation is as follows:
\begin{equation}
    c_{ij}(r)=h_{ij}(r)-\ln\left[h_{ij}(r)+1\right]-\beta u_{ij}(r).
\end{equation}
In the IPY2 closure equation, the bridge function $b_{ij}(r)$ for a like-charge pair is defined as follows \cite{babu1994new}:
\begin{equation}
    b_{ij}(r)=\ln\left[1+\tau_{ij}(r)\right]-\tau_{ij}(r),
\end{equation}
whereas that for an unlike-charge pair is defined as follows:
\begin{equation}
    b_{ij}(r)=\ln\left[2-\left\{1+\tau_{ij}(r)\right\}\exp\{-\tau_{ij}(r)\} \right]\\ .
\end{equation}
The definition of $\tau_{ij}(r)$ is
\begin{equation}\label{eqn:tau}
    \tau_{ij}(r)=h_{ij}(r)-c_{ij}(r)+\tau_{ij}^{\mathrm{long}}(r)\\.
\end{equation}
$\tau_{ij}^{\mathrm{long}}(r)$ in Eq. (\ref{eqn:tau}) is given by
\begin{eqnarray}
    \tau_{ij}^{\mathrm{long}}(r)&=&-\beta \frac{Q_iQ_j}{4\pi\varepsilon_0\varepsilon_{r}r}\left(1-\exp[-\kappa r]\right).
\end{eqnarray}
In this equation, $\kappa$ is Debye screening constant, which is given as
\begin{equation}
    \kappa^2=\frac{\beta}{\varepsilon_0\varepsilon_r}\sum_i \rho_iQ_i^2 .
\end{equation}
This approximation was applied to some molten salt systems \cite{babu1994new}.
The form of $W_{MM}(r)=-k_BT\ln(h_{MM}(r)+1)$ was prepared as follows,
\begin{equation}
    W_{MM}(r)=u_{MM}(r)-k_BT\{t_{MM}(r)+b_{MM}(r)\} .
\end{equation}
Therefore, $W_{MM}(r)$ can be obtained without the iteration process using the abovementioned equation and $t_{MM}(r)$, which is obtained from the third OZ equation (Eq. (\ref{eqn:wavenum})).

\subsection{The IETs and MC Simulation to Calculate Model B}
The radial distribution function $g_{AA}(r) = h_{AA}(r)+1$ is analyzed in Model B. We calculate $g_{AA}(r)$s using the IETs. In this case, we adopt the OZ equation, namely Eq. (\ref{eqn:oz_ij}) for the bulk. Moreover, Model B does not have explicit solvent particles. Thus, $i = C, A, V$ is replaced with $i = C, A$. The OZ equation is solved with the HNC and the IPY2 closures.

MC simulations were performed only for Model B. Because of the Coulomb system, we adopted the Ewald method in the potential energy calculation. The sampling of the Coulomb system was difficult. Therefore, the thermal replica-exchange method was adopted to avoid the trapping at the local minimum. The temperatures of replicas were 298\,K, 320\,K, 340\,K, and 360\,K. The replicas $m$ and $n$ were exchanged using the transition probability\cite{HANSMANN1997parallel,OKABE2001replica}, $P^{acc}_{mn}$:
\begin{equation}
    P^{acc}_{mn}=\mathrm{min}[1,\exp\{-(\beta_m-\beta_n)(U_n-U_m)\}],
\end{equation}
where $U_m$ is the potential energy of replica $m$.

To discuss the electrolyte concentration dependence, the molar concentrations $C_A = $ 0.01\,M, 0.1\,M, 0.3\,M, 0.4305\,M, and 1\,M were examined. According to Ref.\cite{suematsu2018}, the reentrant behavior can be expected in the concentration region for this model. The charge neutrality condition was maintained. Thus, the anion and cation numbers in the MC simulation ($N_A$ and $N_C$) were 60 and 25, respectively ($Q_A=-\,\mathrm{e}$, $Q_C=2.4\,\mathrm{e}$). The basic cell sizes were set according to the concentration.

\section{Results and Discusstions}
\begin{figure}[bt]
    \centering
    \includegraphics[keepaspectratio, scale=0.4]{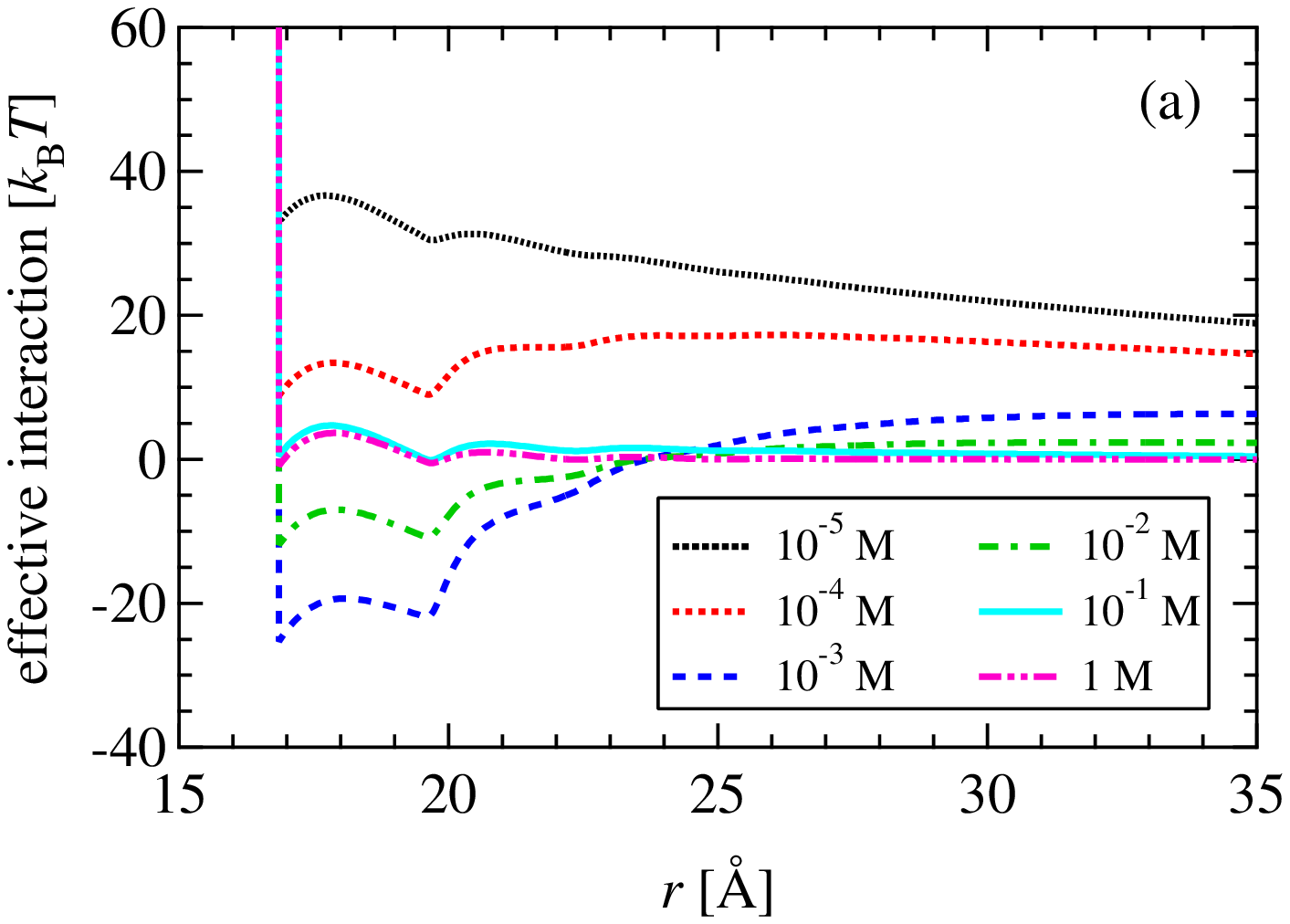}
    \includegraphics[keepaspectratio, scale=0.4]{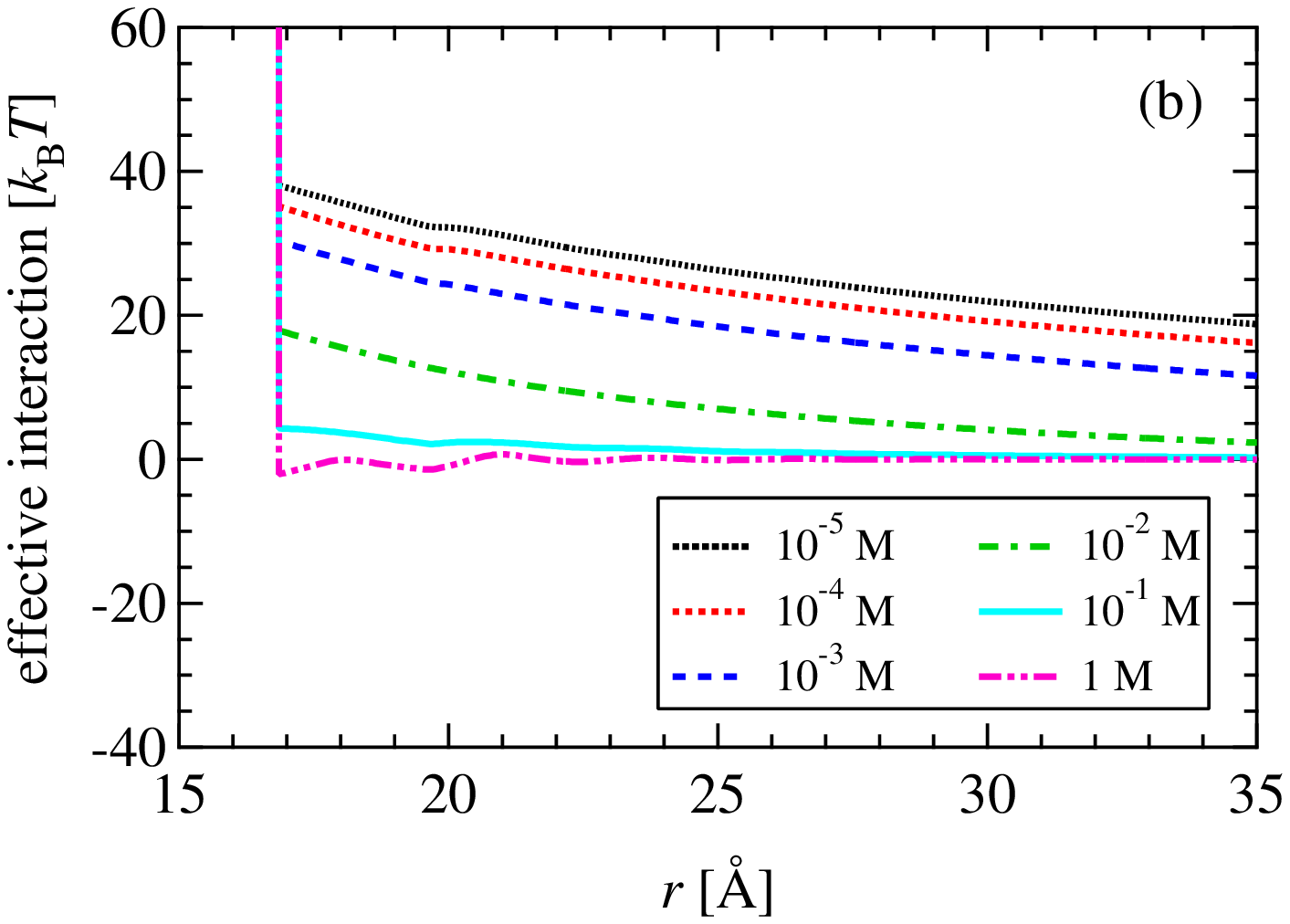}
    \caption{The dependence of the effective interaction between two macroanions on the electrolyte concentration, in Model A. (a) HNC, and (b) IPY2. }
    \label{fig:1}
\end{figure}

The effective interactions between two macroanions are shown in Fig. \ref{fig:1}. Figure \ref{fig:1} (a) presents the effective interactions obtained by the HNC approximation reported in \cite{AkiyamaSakata2011}, and Fig. \ref{fig:1}(b) presents those obtained by the IPY2 approximation. The results will be briefly discussed, as per Ref.\cite{AkiyamaSakata2011}. When the electrolyte concentration was $10^{-5}\,\mathrm{M}$ or $10^{-4}\,\mathrm{M}$, the effective interactions had positive values. When the concentration increased up to $10^{-3}\,\mathrm{M}$, the effective interaction became negative in the vicinity of the contact position. Moreover, when the electrolyte concentration became as high as $10^{-1}\,\mathrm{M}$ or 1\,M, the curves of the effective interaction lay nearly along the zero line, from which it is concluded that the clear attraction disappeared. Therefore, the interaction changed from repulsive to attractive, and the attraction disappeared as the electrolyte concentration increased. Thus, under the HNC, a clear reentrant association was observed between two macroanions.

By contrast, under the IPY2 closure, we obtained the result shown in Fig. \ref{fig:1} (b). When the electrolyte concentration was $10^{-5}\,\mathrm{M}$, a significant repulsion was observed between the two macroanions. As the electrolyte concentration increased, the repulsion gradually disappeared, whereas a clear attraction was not observed in the concentration considered in this study. The results under the IPY2 approximation were qualitatively different from those under the HNC approximation. 
\begin{figure}
    \centering
    \includegraphics[keepaspectratio,scale=0.4]{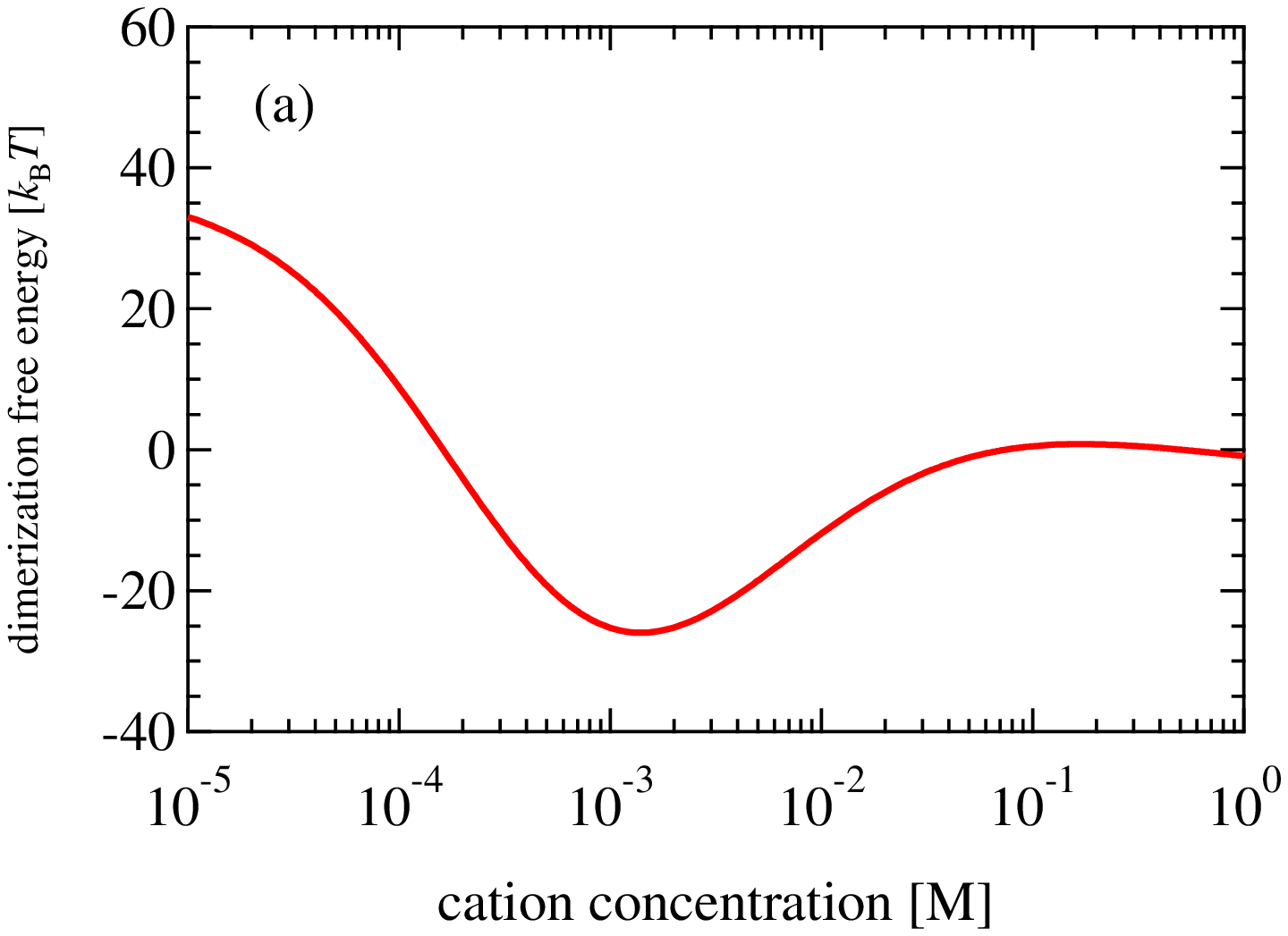}
    \includegraphics[keepaspectratio,scale=0.4]{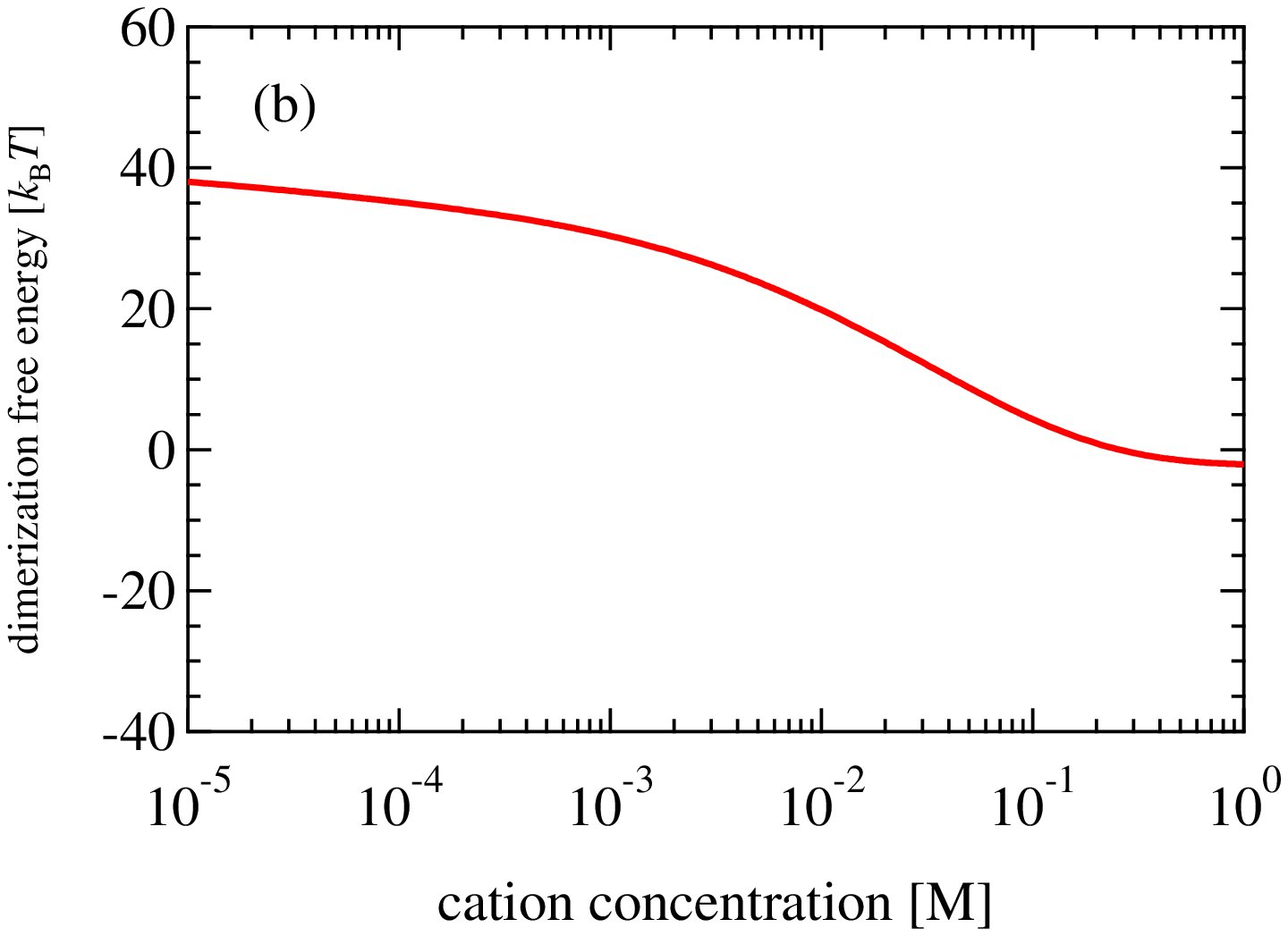}
    \caption{Cation-concentration dependence of the dimerization free energy $\epsilon$ of the effective interaction between macroanions, in Model A. (a) HNC, and (b) IPY2.}
    \label{fig:2}
\end{figure}

Here, we define the dimerization free energy of macroanions $\epsilon$ as the contact value of the effective interaction between macroanions. The electrolyte-concentration dependence of the dimerization free energy $\epsilon$ is shown in Fig. \ref{fig:2}. Figures \ref{fig:2} (a) and (b) correspond to the results obtained by HNC and IPY2, respectively. Under the HNC approximation (Fig. \ref{fig:2} (a)), the dependence of the dimerization free energy $\epsilon$ on the concentration was not monotonic: the dimerization free energy $\epsilon$ decreased with increasing electrolyte concentration, to take a local minimum in the vicinity of $10^{-3}\,\mathrm{M}$, followed by a significant increase. In the low concentration region (i.e., from $10^{-5}\,\mathrm{M}$ to $10^{-4}\,\mathrm{M}$), the dimerization free energy $\epsilon$ was positive, whereas in the middle concentration region (i.e., in the vicinity of $10^{-3}\,\mathrm{M}$), it took a negative value, which was more than 20 times the thermal energy, $k_BT$.  When the concentration became high (i.e., higher than $10^{-1}\,\mathrm{M}$), the dimerization free energy $\epsilon$ was nearly equal to zero. The negative dimerization free energy $\epsilon$ in the vicinity of the local minimum observed in Fig. \ref{fig:2} (a) indicated that the effective interaction between macroanions was attractive, as mentioned above. The curve drawn in Fig. \ref{fig:2} (a) clearly shows the reentrant behavior of the effective interaction under the HNC.
By contrast, the dimerization free energy $\epsilon$ of the effective interaction under the IPY2 approximation decreased monotonically with concentration (Fig. \ref{fig:2} (b)). In the high concentration region, such as the vicinity of $10^0\,\mathrm{M}$,  the dimerization free energy $\epsilon$ was negative (at best twice the thermal energy, $k_BT$). Such a clear attraction as observed under the HNC approximation did not appear when the IPY2 closure was adopted. Moreover, no reentrant behavior was perceived under the IPY2 approximation for the effective interaction between macroanions. These results were in contrast to those of the HNC closure.

The closures showed qualitatively different results: the HNC-OZ theory presented effective attractions between macroanions and the reentrant behavior, whereas the IPY2-OZ theory showed no such behaviors. The effective repulsion between macroanions decreased monotonically with the increasing electrolyte concentration, indicating only screening by the electrolyte in the IPY2-OZ theory.
In the present study, we adopted molecular simulations to study this difference. We surveyed models and thermodynamic conditions for reentrant behavior using the HNC-OZ theory under conditions that could be calculated by molecular simulation \cite{suematsu2018}. We obtained the correlation functions for Model B using the HNC-OZ and IPY2-OZ theories and MC simulation. To discuss the validity of the IETs, we compared the radial distribution functions.

\begin{figure}[bt]
    \centering
    \includegraphics[keepaspectratio,scale=0.4]{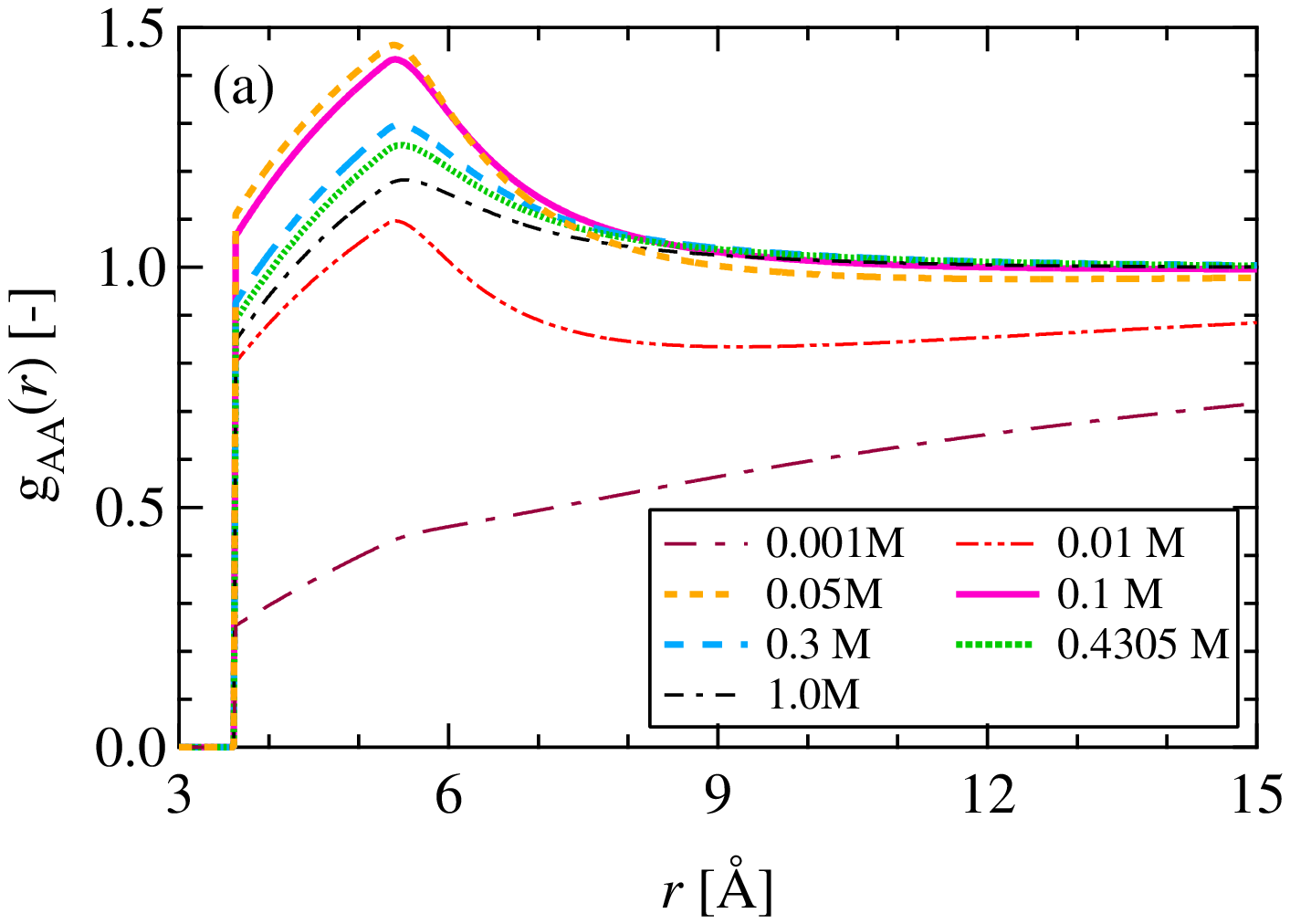}
    \centering
    \includegraphics[keepaspectratio,scale=0.4]{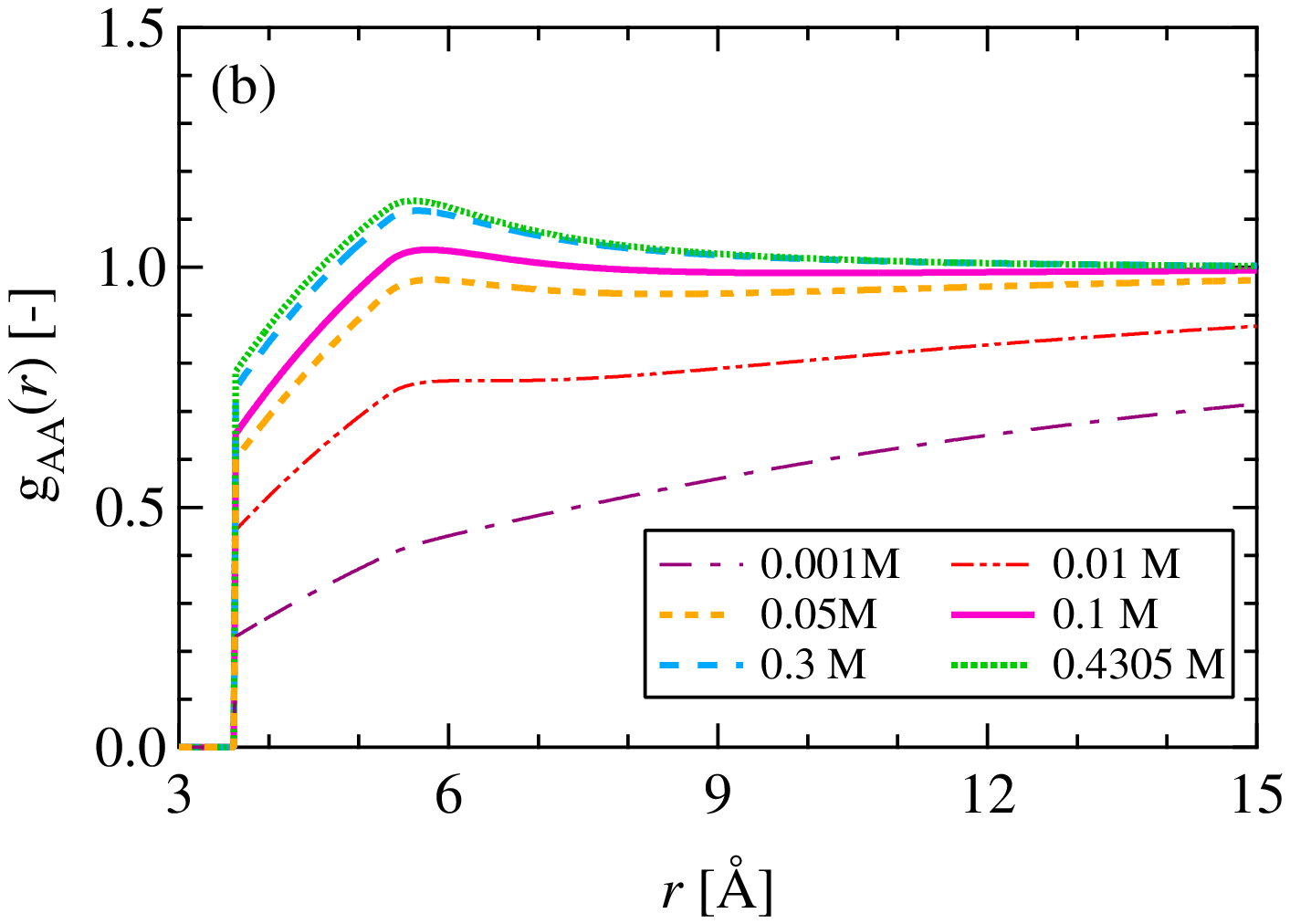}
    \centering
    \includegraphics[keepaspectratio,scale=0.4]{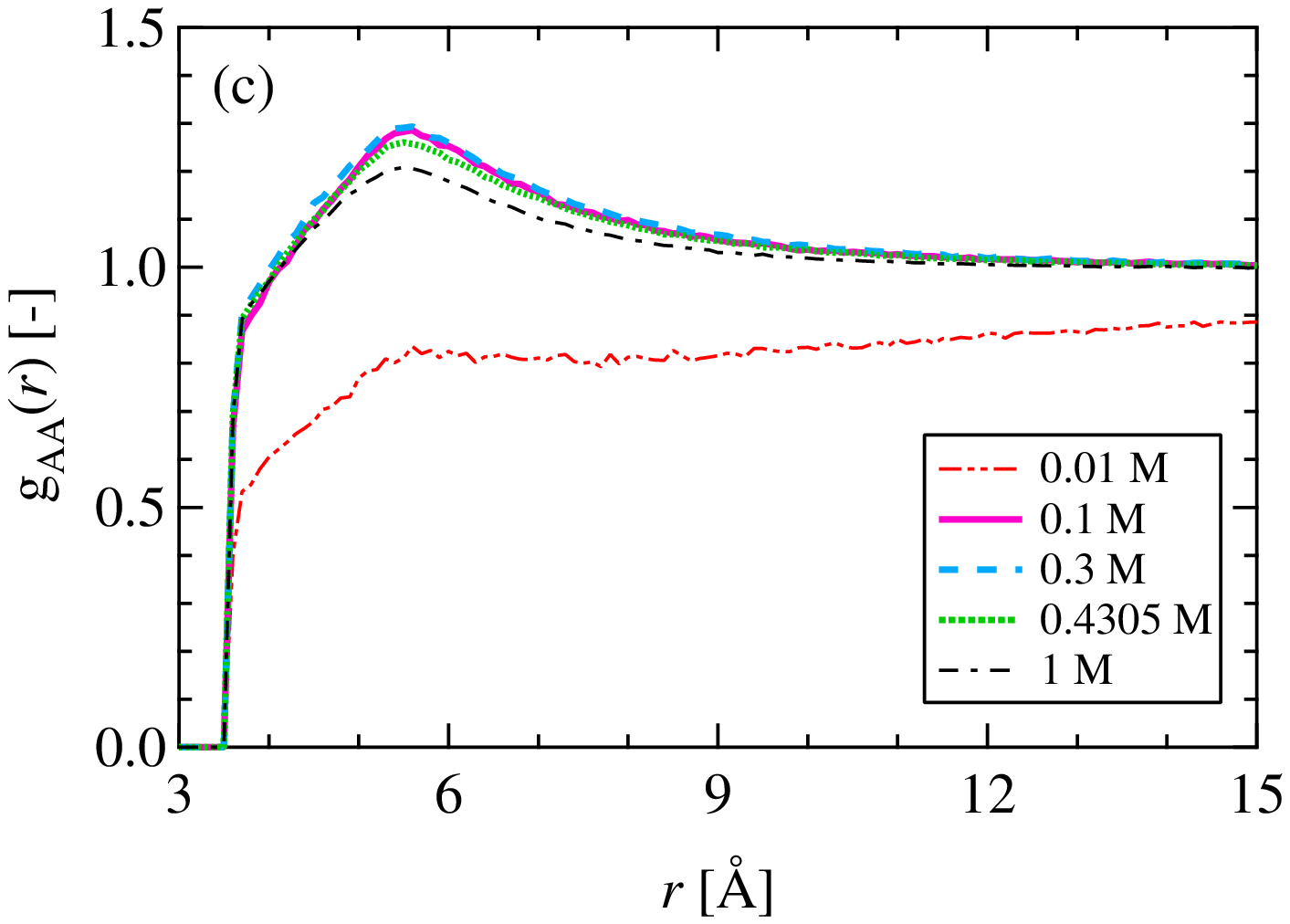}
    \caption{Radial distribution function between anions, $g_{AA}(r)$, in Model B.  (a) HNC, (b) IPY2, and (c) MC.}
    \label{fig:3}
\end{figure}

The radial distribution functions between $Cl^-$ anions $g_{AA}(r)$ are shown in Fig. \ref{fig:3}. Figures \ref{fig:3} (a), (b), and (c) show $g_{AA}(r)$ obtained using the HNC-OZ theory, the IPY2-OZ theory, and the MC simulation, respectively. When the electrolyte concentration was lower than $1.0 \times 10^{-3}\,\mathrm{M}$, all $g_{AA}(r)$s were smaller than 1, meaning that the effective interactions between anions were repulsive. As the concentration increased, $g_{AA}(r)$ became larger. When the concentration was $1.0 \times 10^{-2}\,\mathrm{M}$, only the term $g_{AA}(r)$ for the HNC-OZ theory had values that were larger than 1. Others, namely (b) and (c), followed this behavior, and all $g_{AA}(r)$s had values that were larger than 1 for $1.0 \times 10^{-1}\,\mathrm{M}$ in the vicinity of the contact distance, 5.5\,\AA. This behavior, namely the disappearance of the direct repulsion caused by the increase in electrolyte, is common to all calculation methods when the concentration is lower than $5.0 \times 10^{-2}\,\mathrm{M}$.

\begin{figure}[bt]
    \centering
    \includegraphics[keepaspectratio, scale=0.7]{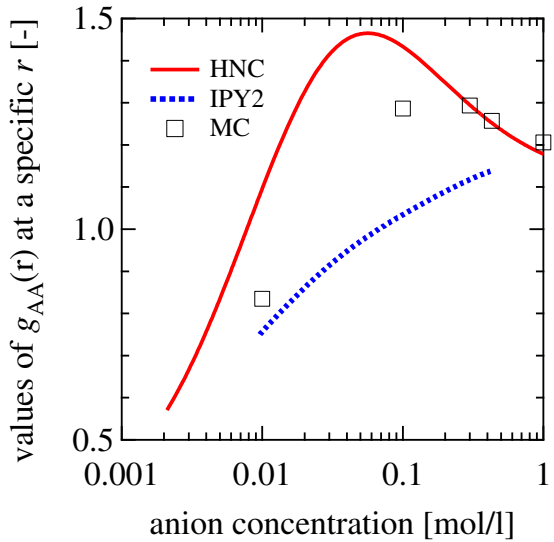}
    \caption{Anion-concentration dependence of the values of $g_{AA}(r)$ at a specified separation in Model B.  The corresponding separations are 5.39\,\AA, 5.61\,\AA, and 5.60\,\AA\, for the HNC, IPY2, and MC, respectively.}
    \label{fig:4}
\end{figure}

Although we found a common behavior under the lower concentration condition, we also found that the HNC approximation emphasized effective attraction in $g_{AA}(r)$. The concentration dependence of the “first peak” value is shown in Fig.\ref{fig:4}. (The definition of the ``first peak'' is given later.) 
The first peak for the HNC approximation was the highest in $g_{AA}(r)$ at $5.0 \times 10^{-2}$\,M. This behavior seems to be due to the well-known short-ranged overcorrelation yielded for many model systems for the HNC approximation\cite{hansen1990theory,perkyns1992site,perkyns1996salting,perkyns2010protein}. In particular, the first peak of the spatial distribution function is emphasized when using the approximation.
By contrast, the values for the IPY2 approximation were closer to those for the MC simulation at $5.0 \times 10^{-2}$\,M. The values for IPY2 were smaller than those of MC. The deviations caused by the IPY2 and HNC approximations were opposite, and the absolute values of the deviations were at the same level. Therefore, there was no superiority between the two closures in this concentration region. The dependence on the electrolyte concentration qualitatively depended on the approximation when the concentration became higher than $5.0 \times 10^{-2}$\,M.

$g_{AA}(r)$s for the HNC-OZ theory and the MC simulation indicated the reentrant behavior. Their first peaks decreased as the concentration increased to certain levels. In Fig. \ref{fig:3}, we also found the highest first peaks for the HNC-OZ theory and MC simulation at $5.0 \times 10^{-2}$\,M and $3.0 \times 10^{-1}$\,M, respectively. Under the high concentration condition, $g_{AA}(r)$ for the HNC-OZ theory became closer to that for the MC simulation. For example, $g_{AA}(r)$ for the HNC-OZ theory was almost the same as that for the MC simulation when the concentration was $3.0 \times 10^{-1}$\,M. By contrast, the value of $g_{AA}(r)$ for the IPY2-OZ theory did not describe reentrant behavior. The concentration dependence was monotonic. Although the concentration dependence for the IPY2-OZ theory had a different qualitative behavior from that for the MC simulation, $g_{AA}(r)$s also got closer to each other at $4.305 \times 10^{-1}$\,M.

The previous paragraph concluded on the similarity and the difference of $g_{AA}(r)$s obtained using three methods with Model B. They are clearly shown in Fig. \ref{fig:4}. However, we must explain how the plots were created because the first peaks of $g_{AA}(r)$ are unclear under the dilute condition. For example, the first peaks are unclear when the concentration is $1.0 \times 10^{-3}$\,M.
We determined the separation distances between two anions for the first peak using the most attractive $g_{AA}(r)$s for each method. The separations of the first peak for the HNC, IPY2, and MC simulation were 5.39\,\AA, 5.61\,\AA, and 5.60\,\AA, respectively. Based on this definition, we obtained the plots in Fig. \ref{fig:4}.
The dots (the IPY2-OZ theory) did not show the reentrant behavior, although others (the HNC-OZ theory (the solid line) and the MC simulation (the squares)) showed it.
At the beginning of this section, we discussed the results for Model A. The difference between the HNC and the IPY2 approximations is similar to that between them in Model B. Our MC simulations suggest that the HNC-OZ theory gives qualitatively reasonable results even when the effective attraction between anions appears. Under this condition, the cations mediate the effective attraction\cite{AkiyamaSakata2011}. Therefore, it seems that the mediate mechanism must be included in the HNC approximation, whereas the mediation effect is excessively emphasized in the approximation.

According to Ref.\cite{babu1994new}, the IPY2-OZ theory gave reasonable correlation functions for highly associating ionic systems such as the 2:2 electrolytes, whereas for monovalent ionic systems the HNC-OZ theory was known to be reasonable \cite{Suematsu2021,Hansen1975,Friedman1981electrolyte}. The model adopted in this study was a mixture of monovalent and multivalent ions, which is an intermediate system between the abovementioned references \cite{babu1994new,Suematsu2021,Hansen1975,Friedman1981electrolyte}. Moreover, this study has dealt with the effective attraction between like-charged particles, which was not considered in the previous studies \cite{babu1994new,Suematsu2021,Hansen1975,Friedman1981electrolyte}. Based on the abovementioned discussion, we can expect the following conclusion. The HNC is better for the mixture of monovalent and multivalent ions than the IPY2. In particular, it seems that the IPY2 approximation includes an insufficient mediation effect. As a result, the approximation did not show the effective attraction and the reentrant behavior.

\section{Conclusion}
We calculated the effective interaction between like-charged particles immersed in an electrolyte solution using two IETs, namely HNC-OZ and IPY2-OZ. Two models were adopted, Models A and B. In Model A, two highly charged macroanions ($-10\,\mathrm{e}$) were immersed in a 1:1 electrolyte solution \cite{AkiyamaSakata2011}. The explicit solvent model was adopted in this model. By contrast, in Model B, the system had small monovalent anions and small multivalent cations. The implicit solvent model was adopted in the latter. 

The electrolyte concentration dependences of the effective interaction were examined. In Model A, the results of the HNC-OZ theory indicated a reentrant behavior. As the concentration increases, the effective interaction turns from repulsive to attractive and returns to repulsive in the reentrant behavior (see Figs. \ref{fig:1} (a) and \ref{fig:2} (a)). By contrast, we did not find the reentrant behavior in the results of the IPY2-OZ theory (see Figs. \ref{fig:1} (b) and \ref{fig:2} (b)). This closure dependence of the reentrant behavior was the same in Model B (Figs. \ref{fig:3} and \ref{fig:4}).

We also performed a MC simulation for Model B. The simulation results qualitatively agreed with those of the HNC-OZ theory. We conclude that the HNC-OZ theory provides adequate results qualitatively. Our previous studies show that the counter-charged ions cause an effective attraction between like-charged particles at the mediation site \cite{AkiyamaSakata2011,FUJIHARA2014,Akiyama2011,suematsu2018,Suematsu2021}. The simulation results suggest that the IPY2 approximation included an insufficient mediation effect. By contrast, the HNC approximation emphasizes the effect\cite{hansen1990theory,perkyns1992site,perkyns1996salting,perkyns2010protein}.

\section*{Acknowledgments}
This work was partly supported by Global Science Campus program of Japan Science and Technology Agency (JST), as well as JST SPRING under Grant Number MJSP2136. This work was also supported by Japan Society for the Promotion of Science (JSPS) KAKENHI Grant Nos. JP21K18604, JP19H01863, JP19K03772, JP18H03673, JP18K03555, and JP16K05659. The computation was performed using Research Center for Computational Science, Okazaki, Japan (Project: 22-IMS-C101 and 22-IMS-C207) and the Research Institute for Information Technology, Kyushu University.

\end{document}